\documentclass[structabstract]{aa}

\usepackage{graphicx}
\usepackage{enumerate}
\usepackage{natbib}
\bibpunct{(}{)}{;}{a}{}{,} 
\usepackage{txfonts}
\usepackage[utf8]{inputenc}
\begin{document}
   \title{Migration of Ca~{\sc ii}~H bright points in the internetwork}

   \author{S.~Jafarzadeh\inst{1,}\thanks{Now at Institute of Theoretical Astrophysics, University of Oslo, Norway.}, R.~H.~Cameron\inst{1}, S.~K.~Solanki\inst{1,2}, A.~Pietarila\inst{3}, 
   			 A.~Feller\inst{1}, A.~Lagg\inst{1}, \and A.~Gandorfer\inst{1}
             }

   \institute{Max-Planck-Institut f\"{u}r Sonnensystemforschung, Max-Planck-Str. 2, 37191 Katlenburg- Lindau, Germany\\
   		 \email{shahin.jafarzadeh@astro.uio.no}
         \and
             School of Space Research, Kyung Hee University, Yongin, Gyeonggi 446-701, Republic of Korea
         \and
             Open Geophysical, Inc., 15021 Katy Freeway, Suite 210, Houston, TX 77094, USA           
             }

   \date{Received 8 November 2013 / Accepted 24 January 2014}

  \abstract
{The migration of magnetic bright point-like features (MBP) in the lower solar atmosphere reflects the dispersal of magnetic flux as well as the horizontal flows of the atmospheric layer they are embedded in.} 
{We analyse trajectories of the proper motion of intrinsically magnetic, isolated internetwork Ca~{\sc ii}~H MBPs (mean lifetime $461\pm9~\mathrm{sec}$) to obtain their diffusivity behaviour.} 
{We use seeing-free high spatial and temporal resolution image sequences of quiet-Sun, disc-centre observations obtained in the Ca~{\sc ii}~H $3968~\mathrm{\AA}$ passband of the Sunrise Filter Imager (SuFI) onboard the Sunrise balloon-borne solar observatory. Small MBPs in the internetwork are automatically tracked. The trajectory of each MBP is then calculated and described by a diffusion index ($\gamma$) and a diffusion coefficient ($D$). We also explore the distribution of the diffusion indices with the help of a Monte Carlo simulation.}
{We find $\gamma=1.69\pm0.08$ and $D=257\pm32~\mathrm{km}^{2}\mathrm{s}^{-1}$ averaged over all MBPs. Trajectories of most MBPs are classified as super-diffusive, i.e. $\gamma>1$, with the determined $\gamma$ being the largest obtained so far to our knowledge. A direct correlation between $D$ and timescale ($\tau$) determined from trajectories of all MBPs is also obtained. We discuss a simple scenario to explain the diffusivity of the observed, relatively short-lived MBPs while they migrate within a small area in a supergranule (i.e. an internetwork area). We show that the scatter in the $\gamma$ values obtained for individual MBPs is due to their limited lifetimes.}
{The super-diffusive MBPs can be described as random walkers (due to granular evolution and intergranular turbulence) superposed on a large systematic (background) velocity, caused by granular, mesogranular, and supergranular flows.}


	\authorrunning{S.~Jafarzadeh~et~al.}
	\maketitle
%

\section{Introduction}

The study of small-scale, magnetic bright point-like features or magnetic bright points (MBPs) in the lower solar atmosphere has gained in interest over the last two decades because they trace $\mathrm{kG}$ magnetic features \citep{Riethmuller2013a}, many of which connect the photosphere with higher layers of the atmosphere~\citep[e.g.][]{Stenflo1989}. Magnetic bright points are among the smallest potentially spatially resolved structures seen in the photosphere~\citep[e.g.][]{Berger2001,Mostl2006,Sanchez-Almeida2010,Riethmuller2010} and chromosphere~\citep[e.g.][]{Rutten1991,Steffens1996,Leenaarts2006,Jafarzadeh2013a}. Their motion is important for the braiding of the magnetic field in the corona; this braiding plays an important role in coronal heating according to~\citet{Parker1972,Parker1983,Parker1988}; cf.~\citet{Gudiksen2002,Gudiksen2005a,Gudiksen2005b,Peter2004}.

Dispersion of the MBPs, i.e. their non-oscillatory motion on the solar surface, is thought to be due to photospheric flows, for example expansion and evolution of granules and supergranules, differential rotation, and meridional flows~\citep{Hagenaar1999}. These motions are commonly described in terms of a diffusion process~\citep[e.g.][]{Leighton1964,Lawrence1993,Dybiec2009,Ribeiro2011} whose efficiency is expressed by a diffusion coefficient ($D$) representing the rate of increase in the area that the MBP diffuses across per unit time. This process can be characterised by the relation $(x-x_{0})^2\propto t^{\gamma}$, where $(x-x_{0})^2$ represents the squared displacement ($sd$) of the tracked MBP between its location $x$ at any given time $t$ and its initial position $x_{0}$; $\gamma$ is normally named the diffusion index. Motions with $\gamma<1$, $\gamma=1$ and $\gamma>1$ are called sub-diffusive, normal-diffusive (random walk) and super-diffusive, respectively. In these regions $sd$ grows more slowly than linear, linearly, or faster than linear with time, respectively~\citep{Dybiec2009}.

Normal-diffusion was historically the first known class of diffusive processes. It characterises a trajectory which consists of successive random steps and is described by the simplest form of diffusion theory~\citep[e.g.][]{Fick1855,Einstein1905,Lemons1997}. Brownian motion is an example of a normal-diffusive process. \citet{Leighton1964} discussed the random walk interpretation of magnetic concentrations in the solar photosphere. He estimated the diffusion coefficient for granules and supergranules to be roughly $10^{4}~\mathrm{km}^{2}\mathrm{s}^{-1}$. He found that this rate is comparable with the dispersal rate of the magnetic regions in the photosphere; and hence concluded that convective flows are responsible for the random walk of the magnetic concentrations~\citep[cf.][]{Jokipii1968,Lawrence1993,Muller1994}.

Sub-diffusive motion of magnetic elements in an active region in the photosphere was first reported by~\citet{Lawrence1993}. Later,~\citet{Cadavid1999} found that although the motion of magnetic network G-band MBPs in the photosphere is random if their lifetimes are larger than $25$ min, MBPs with lifetimes less than $20$ min migrate sub-diffusively. The sub-diffusivity was explained by the trapping of MBPs at stagnation points (i.e. points with zero horizontal velocity; sinks of flow field) in the inter-cellular pattern \citep{Simon1995}. In agreement with \citet{Cadavid1999},~\citet{Hagenaar1999} stated that the diffusion index obtained from tracking magnetic elements may depend on their lifetimes.

There are only a few observational reports of super-diffusion in the lower solar atmosphere. \citet{Berger1998b} found indications of slight super-diffusivity among otherwise normal-diffusive G-band MBPs in network regions. Later, in two high spatial and temporal resolution image sequences,~\citet{Lawrence2001} found a significant number of super-diffusive MBPs between normal- ($\gamma=1$) and Richardson diffusion (i.e. $\gamma=3$). Recently,~\citet{Abramenko2011} observed super-diffusivity of photospheric TiO MBPs in high spatial resolution time series. They reported the presence of super-diffusivity (as the only observed diffusion regime) for both quiet-Sun and active regions. This super-diffusivity was later confirmed by~\citet{Chitta2012} by tracking MBPs observed in wideband H$\alpha$.

Diffusivity of MBPs is thought to be related to the turbulent convection on and below the solar surface~\citep{Nordlund1985}. In addition, it has been shown that stronger magnetic fields and larger magnetic elements result in smaller diffusion coefficients~\citep{Schrijver1989,Schrijver1996}.

In summary, the mobility of small MBPs has been described as dependent on temporal and spatial scales and on the strength of magnetic field.

We present measurements of the motion of small MBPs seen in the high-resolution Ca~{\sc ii}~H images of the SuFI instrument aboard the {\sc Sunrise} balloon-borne solar observatory. The MBPs under study are located in the quiet-Sun internetwork areas sampled at a height corresponding roughly to the temperature minimum. In particular, we track the motion of these small and intrinsically magnetic features (\citealt{Jafarzadeh2013a}, hereafter Paper~I) in time series of filtergrams and calculate their trajectories. We compute the $sd$ for each trajectory and measure the corresponding power exponent (i.e. the diffusion index) as well as the diffusion coefficient.

In Sect.~$2$, we outline the dataset used to produce the time series. Section~$3$ represents the analysis method of the trajectories of the MBPs as well as the diffusion study results. In Sect.~$4$ we explore the role played by the distribution of diffusion indices with the help of a simple model. The concluding remarks are discussed in Sect.~$5$.

\section{Data}
\label{sec-diffData}

For this study we used part of the datasets described in Paper~I. The datasets consist of six time series of intensity images obtained in the Ca~{\sc ii}~H passband (centred at $3968~\mathrm{\AA}$ with FWHM$\approx 1.8~\mathrm{\AA}$) of the {\sc Sunrise} Filter Imager (SuFI;~\citealt{Gandorfer2011}) on board the {\sc Sunrise} balloon-borne solar observatory during its first flight in June 2009~\citep{Solanki2010,Barthol2011,Berkefeld2011}.

\begin{table}[h]
\caption{Log of observations.} 
\label{table:diff-images}
\centering
\begin{tabular}{l c c c}
\hline\hline  \\ [-1.9ex]
Date & Time Interval & No. of & Cadence \\
 & (UT) & frames & \\
 \\ [-1.9ex] 
\hline  \\ [-2.0ex]
   2009 Jun 9 & 00:36 - 00:59 & $117$ & $12~\mathrm{sec}$\\
   2009 Jun 9 & 01:32 - 02:00 & $136$ & $12~\mathrm{sec}$\\
   2009 Jun 11 & 15:22 - 15:44 & $312$ & $4~\mathrm{sec}$\\
   2009 Jun 11 & 20:09 - 20:19 & $78$ & $8~\mathrm{sec}$\\
   2009 Jun 11 & 21:00 - 21:09 & $83$ & $7~\mathrm{sec}$\\
   2009 Jun 13 & 01:46 - 01:59 & $255$ & $3~\mathrm{sec}$\\
\\ [-2.0ex]
\hline     
\end{tabular}
\end{table}

\begin{figure}[h!]
	\centering
	\includegraphics[width=8.5cm]{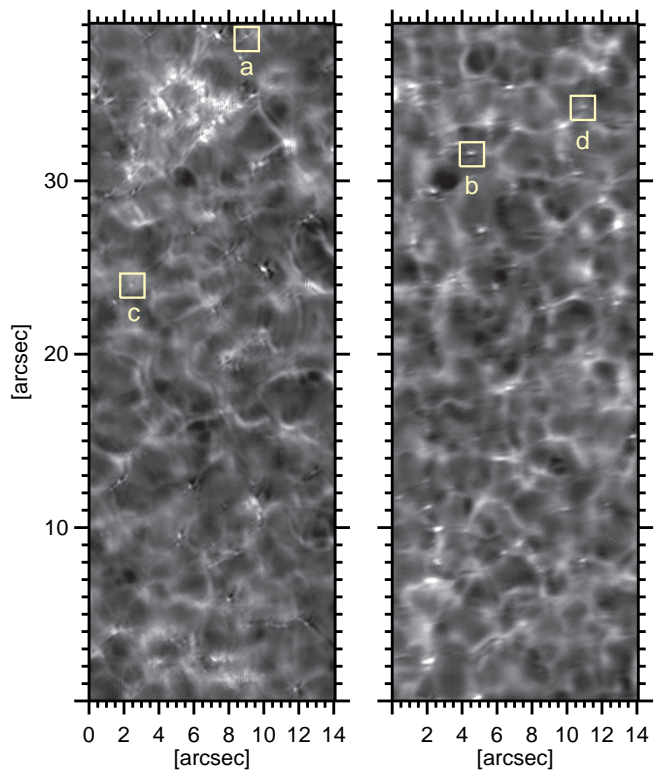}
	\caption{Two examples of SuFI/{\sc Sunrise} Ca~{\sc ii}~H filtergrams. Left: Observed on $9$ June $2009$, at 01:54:43 UT. Right: Obtained on $11$ June $2009$, at 21:05:20 UT. The yellow boxes include selected MBPs whose trajectories are shown in Fig.~\ref{fig:trajectories}.
              }
	\label{fig:diff-obs}
\end{figure}

Table~\ref{table:diff-images} lists the employed image sequences. All images were phase diversity reconstructed employing averaged wavefront errors (see~\citealt{Hirzberger2010,Hirzberger2011}), i.e. they correspond to level 3 data. For all images the field of view is $\approx15\times41~\mathrm{arcsec}^{2}$ ($712\times1972$ pixels) in size, with an image scale of $0.021~\mathrm{arcsec}/ \mathrm{pixel}$. All data refer to quiet regions close to the solar disc centre.

Figure~\ref{fig:diff-obs} illustrates two snapshots from two different time series. We have indicated four MBPs with yellow boxes whose trajectories will be studied here in detail. The left image in Fig.~\ref{fig:diff-obs} contains a small network area in the upper part of the field of view that is excluded from our analysis which concentrates on internetwork MBPs.

\section{Data analysis and results}
\label{sec-diffAnalysisResults}

\begin{figure*}[htp!]
	\centering
	\includegraphics[width=17cm]{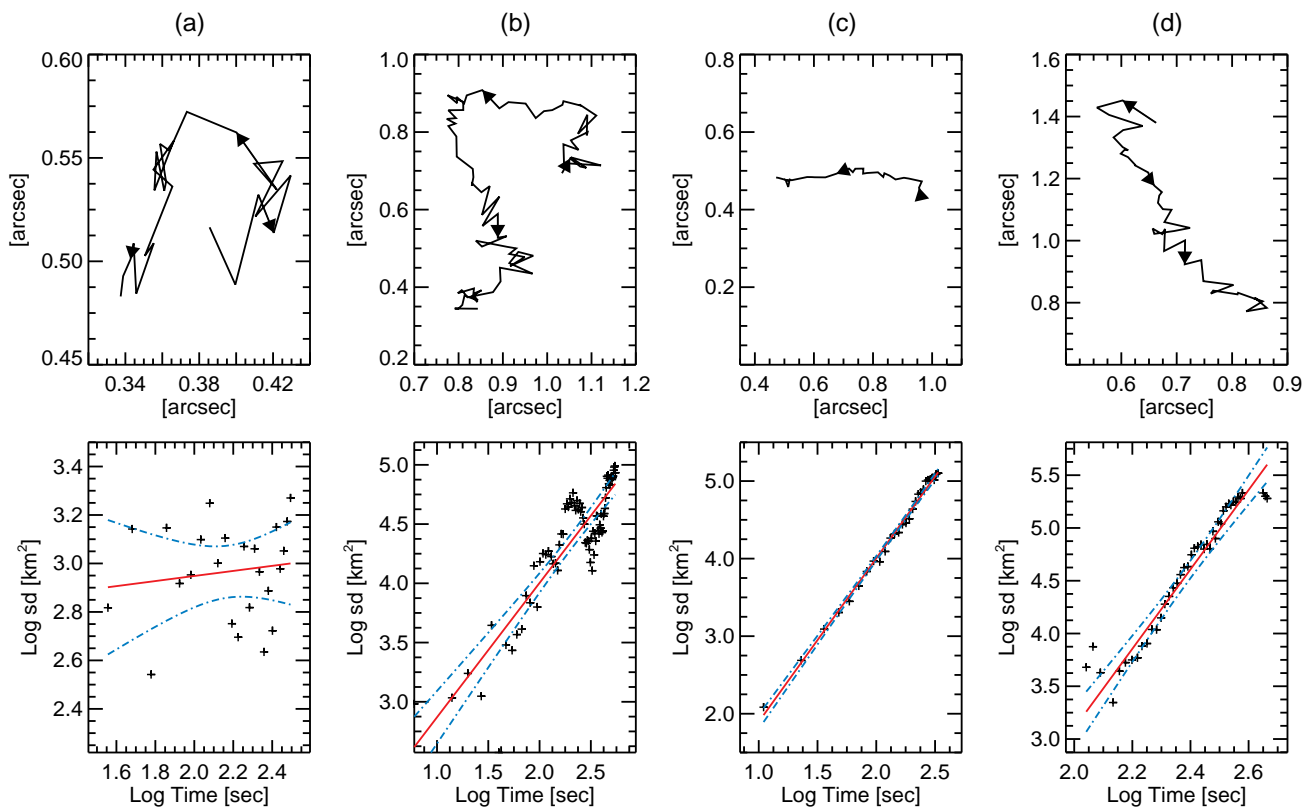}
	\caption{Trajectories (top) and log-log plots of the squared displacement ($sd$) versus time (bottom) of the four MBPs (a-d) identified in Fig.~\ref{fig:diff-obs}, marked at the top of the upper panels. Arrows on the trajectories indicate the direction of the MBPs' motion. The red solid lines are the best linear fits to the $sd$ data points (crosses) vs. time and the blue dot-dashed lines show the $95\%$ confidence bands around the linear fit. The linear fits result in a MBP with $\gamma$=$0.10\pm0.47$ (a), a slope (or $\gamma$) of $1.13\pm0.18$ (b), a MBP with $\gamma$=$2.10\pm0.11$ (c), and a MBP with a high power exponent $\gamma$ of $3.76\pm0.57$ (d). The uncertainties are computed from the $95\%$ confidence bands.
              }
	\label{fig:trajectories}
\end{figure*}

We analyse trajectories of the Ca~{\sc ii}~H MBPs studied in Paper~I. According to the definition introduced in Paper~I, a Ca~{\sc ii}~H MBP is an isolated (no merging, no splitting, and no fine-structure resolved) bright, point-like feature in the height range sampled by the SuFI Ca~{\sc ii}~H $3968~\mathrm{\AA}$ passband (see \citealt{Jafarzadeh2013a} for the relevant contribution function). The MBPs are included in this study if they (1) are located in an internetwork area around the disc centre, (2) are intrinsically magnetic (for those cases for which we have co-spatial \& nearly co-temporal magnetograms), (3) have a lifetime longer than $80~\mathrm{sec}$, (4) have a roughly circular shape with a diameter smaller than $0.3~\mathrm{arcsec}$, and (5) are neither due to acoustic waves nor associated with ``reversed granulation" (cf.~\citealt{Rutten2004}).

The identified (selected) MBPs are tracked using the algorithm described in Paper~I. Furthermore, the trajectories of the MBPs are computed and the modes of motion for each trajectory are investigated.

\subsection{Trajectories}
\label{subsec:trajectories}

The trajectory of MBPs can be reconstructed by linking the string of $x$ and $y$ positions marking the MBPs' locations in each frame. These MBPs have a mean lifetime of $673\pm9~\mathrm{sec}$, when measured only for those MBPs whose birth and death times are observable within the course of the time series and inside the images field of view~\citep{Jafarzadeh2013a}. The observed lifetime was found to be $461\pm9~\mathrm{sec}$ on average when all MBPs are considered. The shortest and longest lifetimes are $167\pm8~\mathrm{sec}$ and $1321\pm13~\mathrm{sec}$, respectively.

Four examples of trajectories of {\sc Sunrise} Ca~{\sc ii}~H MBPs are illustrated in Fig.~\ref{fig:trajectories} (upper panels).

Linking the MBP positions into a trajectory is only possible if the MBPs' displacement in two consecutive frames is sufficiently smaller than the typical distance between the detected MBPs. This condition ensures that interacting MBPs or those that show apparent merging or splitting are excluded. This criterion may cause us to miss the fastest moving MBPs. Since a particular MBP may disappear and reappear again in a short time interval, such as the so-called persistent flashers~\citep{Brandt1992,Jafarzadeh2013a}. Therefore, these possible absent times were considered in the tracking algorithm, while a final visual inspection guaranteed the exclusion of interacting MBPs in these special cases. For details on the locating and tracking procedures of the MBPs, we refer the reader to~\citet{Jafarzadeh2013a}.

\subsection{Diffusion processes}
\label{subsec:diffusion}

The trajectories of MBPs, $\textbf{r}(t)$, for a diffusive process can be parametrised by their self-diffusion coefficients $D$ based on the Einstein-Smoluchowski equation~\citep{Crocker1996},

\begin{equation}
	\left \langle \left | \mathbf{r}\left ( t_{0}+t  \right )-\mathbf{r}\left ( t_{0} \right ) \right |^{2} \right \rangle=2dD\tau\,,
	\label{equ:einstein}
\end{equation}

\noindent
where $\tau$ is the elapsed time and $d$ indicates the trajectory's dimension. The term $\mathbf{r}\left(t_{0}+t\right)-\mathbf{r}\left(t_{0}\right)$ gives the displacement of the MBP at a given time $t_{0}+t$, from its first observed location at time $t_{0}$. In its original definition, the left-hand side of the above equation is the mean of the squared displacement, whose value is computed by averaging all squared displacements ($sd$) in the system, i.e. all squared displacements over all MBPs. However, this averaging can result in the mixing of different diffusive processes and is not ideal for studying different motion types that the MBPs may have~\citep{Dybiec2009}.

In its general form, diffusion is characterised by scaling of the variance of positions or alternatively the $sd$ with time~\citep{Ribeiro2011},

\begin{equation}
	sd\left ( \tau \right ) = C \tau^{\gamma }\,,
	\label{equ:msd}
\end{equation}
\noindent 
where $C$ is the constant of proportionality.
The power exponent $\gamma$ (i.e. diffusion index) is the scaling factor of the $sd$ and, by definition, can be used to classify distinct diffusion regimes as,

\begin{equation}
       \left\{\begin{array}{ll}
             \gamma < 1
              , & \quad \mbox{Sub-diffusive~process}\\
              \gamma = 1
              , & \quad \mbox{Normal-diffusive~process~(Random~walk)}\\
              \gamma > 1, & \quad \mbox{Super-diffusive~process}
       \end{array}\right.
\end{equation}    
\noindent
Since diffusive processes in which the $sd$ grows linearly with time ($\gamma=1$) are called normal, deviations from linearity result in anomalous diffusion. Trajectories for which the $sd$ grows more slowly or more quickly than linearly with time are said to lie in the sub- or super-diffusion regimes, respectively. Diffusive processes with $\gamma =2$ and $\gamma =3$ are often referred to as ballistic and Richardson diffusions, respectively.

In order to compute the $\gamma$, we calculate the $sd$ of each MBP in each image (i.e. at every time-step) from its observed position in the first image it is present in. The diffusion index of the MBP can be conveniently captured by the slope of $sd(\tau)$ on a log-log scale. The standard deviation of the slope is also computed within $95\%$ confidence intervals. Therefore, the true value of the slope and hence of $\gamma$ lies within this confidence interval with a $95\%$ probability.

Once $\gamma$ is known, the diffusion coefficient $D$ is calculated from the constant of proportionality $C$ in Eq.~\ref{equ:msd} for each MBP separately.
Following \citet{Abramenko2011}, the diffusion coefficient, $D$, representing the rate of area in unit time that a MBP moves across, is estimated as the coefficient of turbulent diffusion described by \citet{Monin1975} as a function of timescale,
\begin{equation}
D(\tau ) = \frac{1}{2d}\frac{\mathrm{d} }{\mathrm{d} \tau}(sd(\tau ))\,.
	\label{equ:diffcoeff}
\end{equation}
\noindent
From Eq.~\ref{equ:msd}, it follows that
\begin{equation}
D(\tau ) = \frac{C\gamma }{2d}\tau ^{\gamma -1}\,,
	\label{equ:diffcoeff2}
\end{equation}
\noindent
where $d=2$ for our $2D$ trajectories.

In practice, $C$ can be computed from the constant term of the linear equation obtained from the least-squares fit to the log-log plot of the $sd(\tau)$ for each MBP (i.e. the lower panels of Fig.~\ref{fig:trajectories}; $C=10^{y_{intercept}}$), $\gamma$ is the slope of the fit and $\tau$ represents the MBP's lifetime.

Figure~\ref{fig:trajectories} displays the trajectory and $sd(\tau)$ plots of four selected MBPs indicated as (a)-(d) in Fig.~\ref{fig:diff-obs}, marked at the top of the upper panels in Fig.~\ref{fig:trajectories}. The MBP (a) in Fig.~\ref{fig:trajectories} represents one of the few MBPs (lifetime $312\pm13~\mathrm{sec}$) with a small $\gamma=0.10\pm0.47$. Although this is not the best example in our data, we show its trajectory (top panel) and its log-log plot (bottom panel) here, because it was the only MBP with $\gamma<1$ in the selected frames shown in Fig.~\ref{fig:diff-obs}. The $95\%$ confidence intervals are plotted as the confidence bands (blue dot-dashed lines) around the linear fits (red solid line) to the data points. As can be seen in the upper-left panel, the MBP moves over a small distance and in fact, it stays almost at the same position for some time before making a short jump to the next location. It moves by less than $0.1~\mathrm{arcsec}$ from its initial position, i.e. by less than its own width (size) of $0.16~\mathrm{arcsec}$. The MBP (b) in Figs.~\ref{fig:diff-obs} and \ref{fig:trajectories}, whose lifetime is $548\pm8~\mathrm{sec}$, provides an example where the slope of the linear fit (the red solid line in the bottom panel), $\gamma=1.13\pm0.18$, is consistent with normal-diffusion. This particular MBP displays a peculiar behaviour. The MBP first tends to move quickly from its initial position. Later it changes its general direction and comes closer to its initial coordinates again. Had it disappeared after $300~\mathrm{sec}$, for example, a larger $\gamma$ would have been obtained. This hints that realisation noise, caused by the relatively short lifetimes of the MBPs, may result in a different $\gamma$. We will investigate this effect using a simple model in Sect.~$4$. The $sd(\tau)$ plots of the MBPs (c) and (d) in Figs.~\ref{fig:diff-obs} and \ref{fig:trajectories} (lifetimes $336\pm13~\mathrm{sec}$ and $461\pm8~\mathrm{sec}$) result in $\gamma=2.10\pm0.11$ and $\gamma=3.76\pm0.57$, respectively. Although the two trajectories look rather similar in the upper panels of Fig.~\ref{fig:trajectories}, they are different in the sense that the trajectory of MBP (d) displays a larger random component. The larger $\gamma$ for example (d) comes from the fact that it gets accelerated, i.e. on average it moves faster with time.

\begin{figure}[tp!]
	\centering
	\includegraphics[width=8.2cm]{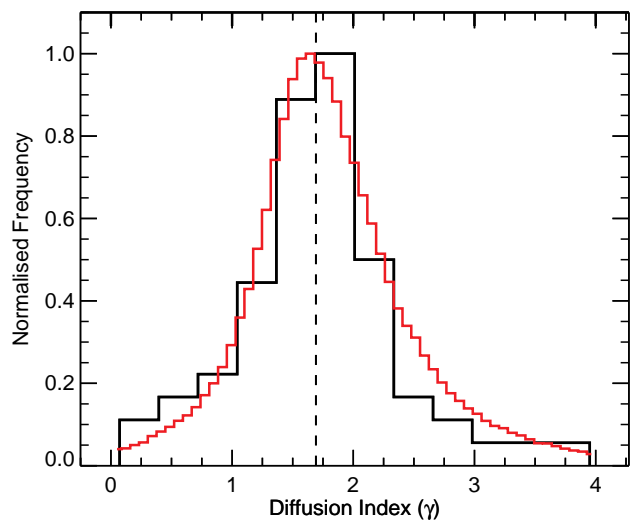}
	\caption{Distribution of the diffusion index $\gamma$ obtained from the {\sc Sunrise}/SuFI data (black histogram), normalised to unity maximum value. The vertical dashed line indicates the mean value of $\gamma$. The red histogram shows the result of a Monte Carlo simulation for the observed MBP (see Sect.~\ref{sec:diff-discussion} for details).
              }
	\label{fig:stat_gamma}
\end{figure}

\subsubsection{Statistics}

Figure~\ref{fig:stat_gamma} shows the statistical distribution of the diffusion indices $\gamma$ of all $103$ MBPs in the analysed  {\sc Sunrise}/SuFI data (black histogram). The histogram is peaked close to $\gamma=2$ and has a mean value of $1.69$ (indicated by the vertical dashed line), with the distribution's standard deviation equal to $0.80$. Eighty-eight percent of these MBPs have $\gamma>1$ and only $12\%$ exhibit $\gamma\leq1$.

The red histogram illustrates the statistics of diffusion indices resulting from a Monte Carlo simulation. We describe and discuss this simulation in Sect.~\ref{sec:MonteCarlo}.

\subsubsection{Diffusion coefficients}
\label{sec:diffusion_coefficients}

Table~\ref{table:diffusion} summarises the computed diffusion coefficients, $D$, for the MBPs a-d (in Figs.~\ref{fig:diff-obs} and \ref{fig:trajectories}) along with their lifetimes and diffusion indices, $\gamma$; $D$ increases with increasing $\gamma$ for these four examples.

\begin{table}[h]
\caption{Diffusion parameters of four example MBPs.}
\label{table:diffusion}
\centering
\begin{tabular}{l c c c}
\hline\hline \\ [-1.9ex] 
MBP & Lifetime & $\gamma$\tablefootmark{$*$} & $D$\tablefootmark{$**$} \\
 & [s] &  & [$\mathrm{km}^{2}\mathrm{s}^{-1}$] \\
\\ [-1.9ex] 
\hline \\ [-1.9ex] 
   a & $312\pm13$ & $0.10\pm0.47$ & $0.08\pm0.01$\\
   b & $548\pm8$ & $1.13\pm0.18$ & $35.8\pm1.2$\\
   c & $336\pm13$ & $2.10\pm0.11$ & $203.6\pm14.7$\\
   d & $461\pm8$ & $3.76\pm0.57$ & $811.7\pm35.6$\\
\\ [-2.0ex]
\hline
\end{tabular}
\tablefoot{
\tablefoottext{$*$}{Diffusion index.}
\tablefoottext{$**$}{Diffusion coefficient.}
}
\end{table}

\begin{figure}[htp!]
	\centering
	\includegraphics[width=8.5cm]{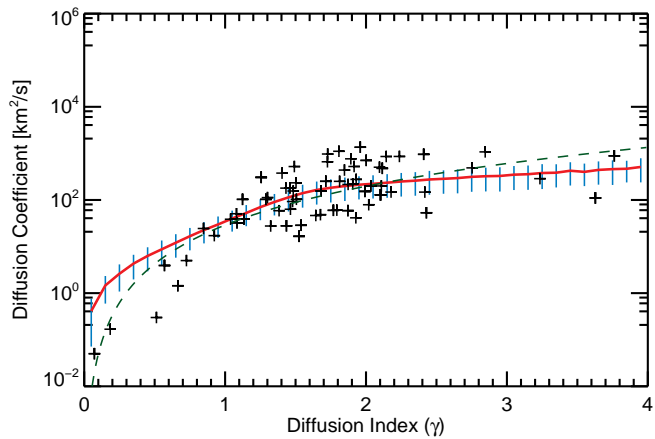}
	\caption{Log-linear plot of diffusion coefficient, $D$, versus diffusion index, $\gamma$. Black crosses indicate the diffusion parameters calculated from trajectories of the {\sc Sunrise} Ca~{\sc ii}~H MBPs. The green, dashed line shows a power-law fit to the data points (see main text). The red solid curve is obtained from a Monte Carlo simulation. Error bars to the simulated values are depicted by blue vertical lines (see Sect.~\ref{sec:MonteCarlo} for details).
              }
	\label{fig:gamma_D}
\end{figure}

The computed $D$ obtained from all {\sc Sunrise}/SuFI Ca~{\sc ii}~H MBPs are plotted as a function of $\gamma$ in a log-linear plot in Fig.~\ref{fig:gamma_D} (black crosses). A power-law fit to the data (i.e. $\gamma = 28.9 \; D^{2.8}$) is overplotted as a green dashed line.

The red line is obtained from the same Monte Carlo simulation as the red histogram in Fig.~\ref{fig:stat_gamma}. In Fig.~\ref{fig:gamma_D} the simulated line has been shifted upward to the observed trend (discussed in Sect.~\ref{sec:MonteCarlo}). The blue vertical lines indicate the error bars of the simulated trend. A mean value of $D=257\pm32~\mathrm{km}^{2}\mathrm{s}^{-1}$ was obtained from the dispersal of the {\sc Sunrise} (internetwork) Ca~{\sc ii}~H MBPs.

\begin{figure}[htp!]
	\centering
	\includegraphics[width=8.5cm]{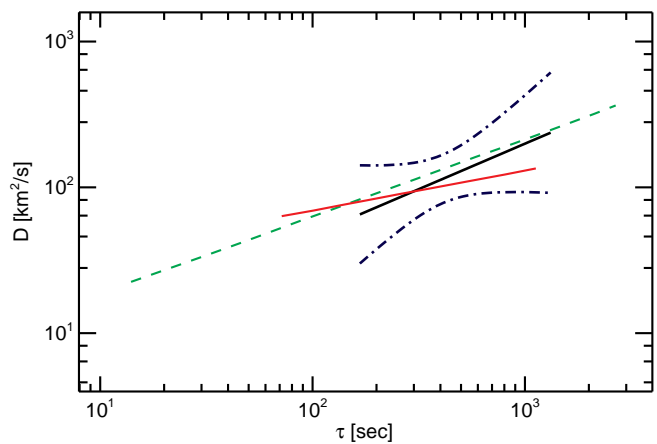}
	\caption{Log-log plot of diffusion coefficients ($D$) as a function of timescale ($\tau$). Black solid line: A linear fit to the data points determined from trajectories of the {\sc Sunrise} Ca~{\sc ii}~H MBPs. The $95\%$ confidence bands to the fit are overplotted as dot-dashed lines. For comparison, similar plots (for quiet-Sun regions) from \citet{Lawrence2001} (red solid line) and \citet{Abramenko2011} (green dashed line) are overlaid.
              }
	\label{fig:t_D}
\end{figure}

Plotted in Fig.~\ref{fig:t_D} is $D(\tau)$ on a log-log scale. The black solid line shows the best linear fit to the data points obtained from the {\sc Sunrise} Ca~{\sc ii}~H MBPs. We observe a direct correlation between the $D$ and timescale ($\tau$) in agreement with the results on super-diffusive internetwork MBPs in the quiet Sun found by \citet{Lawrence2001} (red solid line) and \citet{Abramenko2011} (green dashed line).
The slopes of the linear trends on the log-log plot, determined from our results and those of \citet{Abramenko2011} and \citet{Lawrence2001}, represent power-law exponents of $0.63$, $0.53$, and $0.28$, respectively. Our larger power-law exponent means smaller $D$ in short timescales compared to the other two studies. This may be due to overestimation of short-term motions as a result of seeing in the ground-based data, while on longer timescales seeing effects average out. We also found a direct relationship between $D$ and displacement ($\sqrt{sd}$) values (not shown), similar to that of \citet{Abramenko2011}.

\section{Modelling}
\label{sec:diff-discussion}

The histogram of the diffusion index (i.e. Fig.~\ref{fig:stat_gamma}) is peaked close to $\gamma=2$, with wings reaching down to $\gamma<1$ and up to $\gamma>3$.  Furthermore, more than one third ($39\%$) of the $\gamma$ computed for {\sc Sunrise}/SuFI MBPs are around $\gamma=2$ (within their uncertainties). The fact that our longest lived MBP has a lifetime of $22~\mathrm{min}$ (with each MBP being observed for $\approx8~\mathrm{min}$ on average) leads us to speculate that the broad wings of the distribution are due to the relatively small number of individual frames (steps) at which we have observed each MBP, i.e. the width of the distribution is due to realisation noise caused by relatively short lifetimes of the MBPs. Clearly, the $sd$ can be fit more precisely when a longer MBP motion is followed. This can be clarified by making a simple model of the horizontal velocity of a MBP. We note that we are dealing with internetwork MBPs, which can lie anywhere in a supergranule (SG).

\subsection{Migration scenario}
\label{sec:scenario}

The MBPs are located in a complex turbulent medium. Correspondingly, the observed horizontal velocity of a MBP can be broken up into several components. In general, a variety of motions advect the MBPs, acting on different time and spatial scales. These motions range from turbulence in intergranular lanes (acting on the order of a few seconds), via the expansion and motion of granules (on the order of a few minutes), meso- and supergranular flows (on the order of one hour and about one day, respectively) to differential rotation and meridional flow. Given the relatively small field of view of all analysed timeseries and their locations close to disc centre, the influence of the meridional circulation and differential rotation can be neglected.

The relatively short lifetimes of the MBPs under study implies that they move only over a fraction of a meso-/supergranule in the course of their average lifetime of $\approx 8~\mathrm{min}$. This in turn implies that the short-lived small-scale motions (i.e. intergranular turbulence and granular evolution) are primarily responsible for the random walk components of the MBPs' motion. Conversely, we expect that the the meso- and supergranular flows as well as motions imparted by constantly expanding, contracting, and splitting granules on passively advected MBPs can be considered to be systematic.

We note that granules can contribute in a way to both random ($v_{r}$) and systematic ($v_{s}$) components of the MBPs' horizontal velocity, since the speeds imparted on MBPs due to granular evolution (slowly) change in the course of a MBP's lifetime.

In addition, it is worth mentioning that a part of the motion of the Ca~{\sc ii}~H MBPs can be due to kink waves excited at these magnetic elements. However, we do not search for any periodicity in the motion of the MBPs so that the effects of any kink waves running along the field lines underlying the MBPs are assigned to the random motion.

\subsection{Monte Carlo simulation}
\label{sec:MonteCarlo}

Assuming an explanation as proposed in Sect.~\ref{sec:scenario}, we explore the statistical distribution of the diffusion index $\gamma$ (shown in Fig.~\ref{fig:stat_gamma}) and its relationship with $D$ (plotted in Fig.~\ref{fig:gamma_D}) obtained from the {\sc Sunrise}/SuFI data with the help of a Monte Carlo simulation.

We allow $30000$ randomly generated ``MBPs" to move along the perpendicular $x$ and $y$ axes in $2D$ space and in time $t$. The $x$ and $y$ axes represent the directions of the random motions of a MBP, while the systematic direction is identified with the $x$ axis. As in Sect.~\ref{sec:scenario} the latter can be taken to represent mainly mesogranular, supergranular and granular flows. The random motions are thought to be largely associated with granular and turbulent intergranular flows (on short timescales). The random walk is modelled assuming a discrete time-step of $\Delta t$ (coherence time), over which the velocity is assumed to be constant. The $\Delta t$ and the ratio of the random to systematic velocities ($v_{r}/v_{s}$) are treated as initial free parameters of the simulation.

We performed the Monte Carlo simulations for MBPs with lifetimes between $160$ and $1320~\mathrm{sec}$; $30000$ realisations were analysed for each lifetime.
For each run of the simulation (for a particular lifetime), the trajectories of the $30000$ simulated MBPs were determined in the same way as for the observed MBPs, described in Sect.~\ref{subsec:trajectories}, and are also analysed to compute their diffusion indices using a method similar to that applied to the {\sc Sunrise}/SuFI Ca~{\sc ii}~H MBPs as described in Sect.~\ref{subsec:diffusion} (i.e. the slope of the log-log plot of the $sd(\tau)$ for each MBP is computed from the simulated MBP trajectory). Therefore, for any set of the two free parameters, we obtain a distribution of diffusion index $\gamma$ for the $30000$ simulated MBPs with a given lifetime.

\begin{figure}[tb!]
	\centering
	\includegraphics[width=8.5cm, trim = 0 0 0 0, clip]{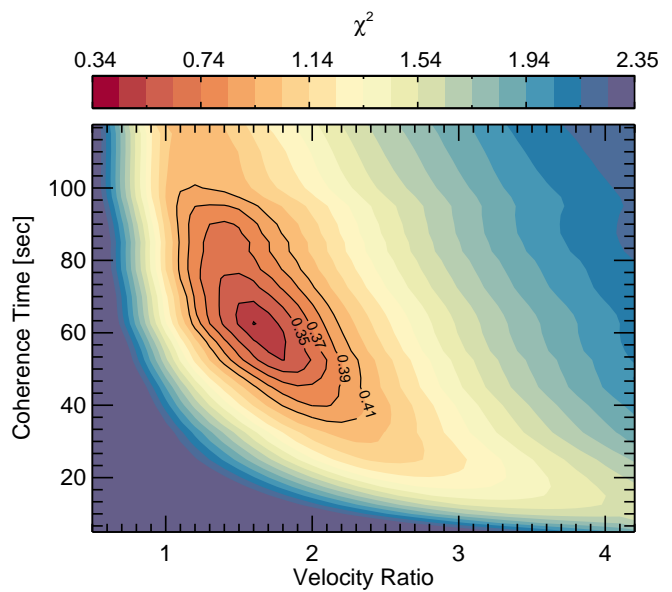}
	\caption{Results of a chi-square ($\chi^{2}$) test for finding the best match between the distributions of the diffusion index from observations and Monte Carlo simulations of the MBPs located in a supergranule's bulk. The contours guide the eye to the best $\chi^{2}$ value, giving the appropriate coherence time $\Delta t$ for the random velocity and the ratio of random-to-systematic velocities $v_{r}/v_{s}$ of the MBPs under study.
              }
	\label{fig:chiStv}
\end{figure}

Integration of the individual simulated histograms corresponding to each lifetime results in a distribution of $\gamma$ calculated for MBPs with different lifetimes, similar to the distribution obtained from the observations. To compare with the observed distribution, the histogram resulting from the simulation is re-binned to match the resolution of the one obtained from the observations. Then, the free parameters are tuned until the best match between these two histograms (i.e. one obtained from the observations and one calculated from the Monte Carlo simulation) occurs. The chi-square, $\chi^{2}$, is computed as a measure of the best match between the two histograms. Since we compare the distributions of the two binned datasets, the quantity $\chi^{2}$ is measured as~\citep{Press2007}
\begin{equation}
	\chi ^{2}=\sum_{i} \frac{\left ( O_{i}-C_{i} \right )^{2}}{O_{i}+C_{i}}\,,
	\label{equ:chiS}
\end{equation}
\noindent
where $O_{i}$ and $C_{i}$ indicate the $i^{th}$ observed and computed (simulated) bin of the histograms.\\
In Fig.~\ref{fig:chiStv} the coherence time $\Delta t$ is plotted as a function of the ratio of velocities $v_{r}/v_{s}$ for all $\chi^{2}$ values. The $\Delta t$ and $v_{r}/v_{s}$ combination giving the smallest $\chi^{2}$ value was found to be $62\pm3~\mathrm{sec}$ and $1.6\pm0.03$. We note that the absolute values of the velocity are not constrained by the histogram of $\gamma$; therefore for simplicity we used $v_{r}=1~\mathrm{km}\, \mathrm{s}^{-1}$. 

The result of the simulated distribution of $\gamma$ based on these parameters (integrated over all simulated distributions with different lifetimes) is plotted in Fig.~\ref{fig:stat_gamma} as a red histogram. A comparison with the black histogram, which represents the distribution of diffusion index obtained from the observations (i.e. from the {\sc Sunrise}/SuFI data), shows that the simulated distribution is stronger than that of the observations in the (far) wing at $\gamma>2.5$, but it is weaker for $\gamma<1$.

In the next step, we determine the absolute values of these random and systematic velocities using the diffusion coefficient $D$. Given the determined ratio of the velocity components as well as the coherence time, we compute $D$ in the same way as for the observed data, described in Sect.~\ref{sec:diffusion_coefficients}. Then, we overplot the computed $D$ versus $\gamma$ for the simulated data on the same log-linear plot as for the observed data. For simplicity, we average the $D$ values in each $\gamma$ bin and overplot its standard deviation as an error bar for each point. The plot of log($D$) versus $\gamma$ in Fig.~\ref{fig:gamma_D} shows similar trends for both simulated and observed data, but originally with an offset in the direction of log($D$).

We have only one free parameter left in the simulation, namely $v_{s}$, with which to shift the log($D$) versus $\gamma$ curve. We note that an arbitrary value for $v_{r}$ had been used so far. Therefore, we tuned $v_{r}$, while keeping $v_{r}/v_{s}$ and $\Delta t$ fixed, until we found the best agreement between the observed and computed plots of log($D$) versus $\gamma$. The output of the simulation giving the best agreement with the observations is overplotted as a red line (along with the error bars; vertical blue lines) on the observed data (black crosses) in Fig.~\ref{fig:gamma_D}. The $v_{r}$ giving the best fit was found to be $1.2\pm0.1~\mathrm{km}\, \mathrm{s}^{-1}$ on average. Consequently, the systematic flow tends to move with a velocity of $0.75\pm0.06~\mathrm{km}\, \mathrm{s}^{-1}$.

\begin{figure}[tb!]
	\centering
	\includegraphics[width=8.5cm, trim = 0 0 0 0, clip]{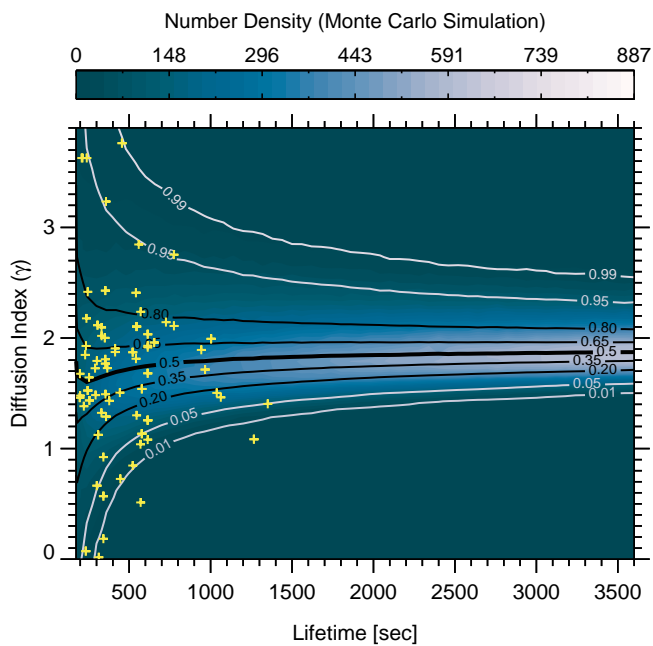}
	\caption{Evolution of the diffusion index histogram with increasing lifetime of the MBPs. The background brightness is a measure of the number density of MBPs with a certain $\gamma$ obtained from a Monte Carlo simulation. The solid lines indicate the percentiles of the distributions (see main text for details). The observed data points from the {\sc Sunrise}/SuFI data are overlaid on the simulated plot and are marked by yellow crosses.
              }
	\label{fig:simulation}
\end{figure}
\noindent  

Finally, we take these best-fit values and do the simulation once again, but for a larger range of lifetimes, i.e. from $80~\mathrm{sec}$, which represents the criterion for minimum lifetime of the MBPs under study, to three times longer than the maximum lifetime of the observed MBPs from the {\sc Sunrise}/SuFI data. A plot of $\gamma$ versus lifetime resulting from this simulation is illustrated in Fig.~\ref{fig:simulation}. The background brightness follows the number density of MBPs with a particular $\gamma$. A vertical cut at a given lifetime indicates the histogram density of the $\gamma$ distribution at the chosen lifetime. The solid lines mark percentiles, i.e. fixed values of the integrals over the histogram density in the vertical direction starting at the bottom. Each yellow cross represents an observed MBP from the {\sc Sunrise}/SuFI data. At the left side of this plot, where most of our observed data points are located, the lifetimes are shorter and correspond to much fewer time-steps than the right side. The left part of the plot displays a wider distribution of $\gamma$, with the histograms having extended tails, similar to the tails of the observed histogram in Fig.~\ref{fig:stat_gamma}. Insufficient sampling on the left part of the diagram enhances the effect of the random velocity, so that an individual MBP can display a $\gamma$ rather far removed from the expectation value.

There are only few data points (as seen in Fig.~\ref{fig:simulation}) that have $\gamma$ values that are highly unlikely according to the simulations, outside the $5\%$ and $95\%$ percentile curves, for example. One source of these outlying MBPs may be the fact that we considered a uniform systematic flow dragging the MBPs with it, whereas in reality the flow speed (caused by the contraction, expansion or explosion of granules as well as the meso- and supergranular flows) is expected to vary. However, the simulated histogram includes almost all of the data points and so can explain most of the observations. Interestingly, the four data points whose locations lie outside the outermost ($1\%$) contour of the simulation in Fig.~\ref{fig:simulation} all have small $\gamma$. One possibility to explain the excess of low $\gamma$ MBPs is that these are lying at the borders of the SGs, where sub-diffusive and random walk MBPs are expected and are due to inflows from opposite directions (from neighbouring SGs).

To summarise, the observations are consistent with the MBPs being random walkers superposed on a systematic flow. Therefore, while the motion of MBPs has a random component due to intergranular turbulence as well as the birth and death of neighbouring granules, the MBPs are transported by a systematic velocity on a larger spatial scale, e.g. due to the constant evolution of long-lived granules, or the contributions of meso- and supergranules to the large-scale velocity field felt by the magnetic element underlying the MBP.

We found a coherence time of $62\pm3~\mathrm{sec}$ which is comparable with a value of $68~\mathrm{sec}$ computed from the given size (length scale) of our features (i.e. $150~\mathrm{km}$ on average) as well as the mean horizontal velocity of the MBPs under study (i.e. $2.2~\mathrm{km}\, \mathrm{s}^{-1}$) reported by \citet{Jafarzadeh2013a}. This implies that on average MBPs move a distance corresponding to their own diameter before being forced to move in another direction.

\section{Discussion and conclusions}
\label{sec-diffConclusions}

We analysed the trajectories of $103$ isolated (i.e. displaying no merging or splitting) internetwork MBPs ($\approx0.2~\mathrm{arcsec}$ in diameter on average;~\citealt{Jafarzadeh2013a}) in the quiet Sun observed in the Ca~{\sc ii}~H $3968~\mathrm{\AA}$ passband of {\sc Sunrise}/SuFI. We performed a diffusion analysis on the trajectories of MBPs to distinguish between MBPs with different types of motions. We did not, however, search for oscillations or wave-like motions.

In order to avoid mixing MBPs in different diffusion regimes and to get a better insight into the character of their proper motion, we performed the diffusion analysis on all individual trajectories separately~\citep{Dybiec2009}.

We used the same MBPs as in Paper~I where they were identified using stringent criteria. This gives us confidence that these MBPs are good tracers of small-scale magnetic elements in the upper layers of the photosphere and the lower chromosphere. These intrinsically magnetic features can be considered to represent the cross-sections of nearly vertical flux tubes \citep{Jafarzadeh2013p} whose positions are influenced by different external forces and are restricted to intergranular lanes. Therefore, although they are not perfect tracers of the horizontal flows, the spatial and temporal scaling of the MBPs' dynamics can still indicate the presence of turbulence in their paths~\citep{Lawrence2001}.

We should note that the relatively small field of view ($\approx15\times41~\mathrm{arcsec}$), short time series (less than $28~\mathrm{min}$), and the restrictive criteria applied to select these small MBPs have limited our sample to a relatively small number of $103$ isolated MBPs (a number density of $0.03~(Mm)^{-2}$; Paper~I). However, selecting only point-like features has allowed us to locate the MBPs more accurately.

\subsection{Diffusion index}
\label{sec-diffConclusions-gamma}

\begin{table*}[!htbp]
\caption{Comparison of the mean values of diffusion index $\gamma$ and diffusion coefficient $D$ of small magnetic elements obtained in this study with some of those in the literature.}
\label{table:compGamma}
\centering                          
\begin{tabular}{l c c c c c c c}       
\hline\hline \\ [-1.8ex] 
Reference & Origin & Telescope/ & Spatial & Feature\tablefootmark{$a$} & Lifetime\tablefootmark{$b,c$} & $\gamma$\tablefootmark{$c$} & $D$\tablefootmark{$b,c$} \\    
 & of data & Spacecraft & resolution &  & [$\mathrm{s}$] &  & [$\mathrm{km}^{2}\mathrm{s}^{-1}$] \\
\\ [-1.9ex] 
\hline \\ [-1.9ex] 
This study & Stratospheric & {\sc Sunrise}/SuFI & 0.$''$14 & IMBP & $461$ & $1.69$ & $257$\\
 & balloon &  &  &  &  &  & \\
\citet{Chitta2012} & Ground & SST/CRISP & - & IMBP & $180-240$ & $1.59$ & ($\approx90$)\\
\citet{Abramenko2011} & Ground & BBSO/NST & 0.$''$11 & IMBP\tablefootmark{$d$} & ($10-2000$) & $1.48$ & ($19-320$)\\
\citet{MansoSainz2011} & Space & Hinode/SOT & 0.$''$32\tablefootmark{$e$} & IMBP & $<900$ & $0.96$ & $195$\\
\citet{Utz2010} & Space & Hinode/SOT & 0.$''$22\tablefootmark{$f$} & IMBP & $150$ & ($\approx1$) & $350$\\
\citet{Chae2008} & Space & Hinode/SOT\tablefootmark{$g$} & - & ME\tablefootmark{$d$} & - & - & $0.87$\\
\citet{Lawrence2001} & Ground & SVST & 0.$''$23 & NMBP & $9-4260$ & $1.13$ & -\\ 
\citet{Cadavid1999} & Ground & SVST & 0.$''$23\tablefootmark{$h$} & NMBP & $18-1320$ & $0.76$ & -\\
 &  &  &  &  & $1500-3450$ & $1.10$ & -\\
 \citet{Hagenaar1999} & Space & SOHO/MDI & 2.$''$3\tablefootmark{$i$} & MFC & $< 1.0\times10^{4}$ & ($\approx1$) & $70-90$\\ 
 &  &  &  &  & $>3.0\times10^{4}$ &  & $220-250$\\
\citet{Berger1998b} & Ground & SVST & $\approx$0.$''$2 & NMBP\tablefootmark{$d$} & ($100-3800$) & $1.34$ & $60$\\
\citet{Lawrence1993} & Ground & BBSO & - & ME\tablefootmark{$d$} & - & $0.92$ & $250$\\
\\ [-1.9ex] 
\hline
\end{tabular}
\tablefoot{
\tablefoottext{$a$}{IMBP: Internetwork Magnetic Bright Point; NMBP: Network MBP; MFC: Magnetic Flux Concentration; ME: Magnetic Element.}
\tablefoottext{$b$}{Mean values; otherwise a range is given whenever it was available.}
\tablefoottext{$c$}{Values in parentheses are not directly reported in the papers, but are computed from the given plots/tables.}
\tablefoottext{$d$}{See their paper for other types of regions.}
\tablefoottext{$e$}{From \citet{Martinez-Gonzalez2009}.}
\tablefoottext{$f$}{From Table~1 of \citet{Abramenko2011}.}
\tablefoottext{$g$}{From SOT magnetograms with the smallest pixel size of $116~\mathrm{km}$.}
\tablefoottext{$h$}{From \citet{Lawrence2001}. \citet{Cadavid1999} reported a much lower resolution element of 0.$''$4.}
\tablefoottext{$i$}{For five-minute averaged and spatially smoothed high-resolution magnetograms.}
}
\end{table*}

We found a mean diffusion index (or diffusion power-law exponent), $\gamma$, of $1.69\pm0.08$ averaged over all MBPs. The $\gamma$ histogram is peaked close to this value and ranges from $\gamma \approx 0$ to $\gamma \approx 4$ (see Fig.~\ref{fig:stat_gamma}). We suspect, however, that this wide range in the $\gamma$ distribution does not reflect the presence of different diffusion regimes (i.e. sub-, normal- or super-diffusive; as described in Sect.~\ref{subsec:diffusion}), but that it is possibly due to realisation noise (around the mean value representing a single diffusion category) due to our relatively short-lived MBPs (mean lifetime $\approx8~\mathrm{min}$). The short lifetime may in turn be due to the fact that we consider only small ($<0.3~\mathrm{arcsec}$ diameter) internetwork MBPs.

We also explored the distribution of the diffusion index with the help of a Monte Carlo simulation (see Sect.~\ref{sec:MonteCarlo}). It demonstrated that the migration of internetwork MBPs is consistent with a random walk (due to intergranular turbulence and granular evolution) superposed on a systematic velocity (caused by granular as well as meso- and supergranular flows). We note that a part of the motion of the Ca~{\sc ii}~H MBPs can be due to kink waves excited in the underlying magnetic elements, whose effects were not separated in our study from that of random motion. The simulation clarified and confirmed that the deduced $\gamma$ values of almost all MBPs are consistent with a single underlying $\gamma$ of $1.69\pm0.08$ and that the large scatter of $\gamma$ values of individual observed MBPs is indeed caused by the short lifetimes of these MBPs and the associated realisation noise. Hence, the dispersal of the studied Ca~{\sc ii}~H MBPs in the quiet Sun is mainly super-diffusive.\\
We note that excluding the network regions in the present study most likely has also excluded most MBPs located at stagnation points and hence, sub-diffusive MBPs may be under-represented in our sample, compared to the entire quiet Sun.\\
Our main result is in rough qualitative agreement with that of~\citet{Lawrence2001}, who found $\gamma=1.13\pm0.01$ (on average) for MBPs, using high spatial and temporal resolution images in a network region of enhanced magnetic activity. However, the features we studied display a much stronger super-diffusive behaviour. Super-diffusivity has also been reported by~\citet{Abramenko2011} and \citet{Chitta2012} in recent investigations. They deduced values of the diffusion index in a given type of region by averaging the displacements over all MBPs. \citet{Abramenko2011} obtained $\gamma=1.48$ for quiet Sun, $\gamma=1.53$ for plage, and $\gamma=1.67$ for coronal hole areas, while \citet{Chitta2012} reported a diffusion index of $1.59$ for their relatively short-lived MBPs (with a mean lifetime of about $3-4~\mathrm{min}$) observed in quiet-Sun regions. The mean $\gamma$ for the quiet Sun obtained in both studies are roughly consistent with the average $\gamma=1.69\pm0.08$ that we found in the present work, although their values are slightly smaller. This may have to do with the fact that earlier $\gamma$ values were obtained from ground-based observations and could be affected by differential seeing-induced deformations (introducing artificial turbulences), while we determined $\langle \gamma \rangle$ using the {\sc Sunrise}/SuFI data unaffected by seeing.

\citet{Lepreti2012} used the same datasets as \citet{Abramenko2011} to determine $\gamma$ (for timescales $<400~\mathrm{sec}$) for pair dispersion of MBPs (i.e. from measuring the mean-square separation of pairs of MBPs). They found the same $\gamma$ ($\approx1.48$) for MBP pairs observed in all the three regions (i.e. quiet-Sun, plage, and coronal hole areas), from which they interpreted the diffusivity properties as the results of the local correlations in the turbulence's inertial range. In addition, \citet{Lepreti2012} concluded that the diffusivity of individual MBPs studied by \citet{Abramenko2011} (that differs in the three regions) depends on the detailed structure of the flows.

\citet{MansoSainz2011} found in contrast to other recent studies a mean value of $0.96$ (nearly corresponding to normal-diffusion) averaged over a wide distribution of $\gamma$ for magnetic internetwork elements. We suspect that such a small value is due to their criterion of tracking only the footpoints of small-scale magnetic loops. The motion of this freshly emerged field is only partly driven by flows at the solar surface, while to a significant extent it also reflects the dynamics and subsurface structure of the emerging field. Hence their results may not be directly comparable with ours.

Some published values of $\gamma$, of the diffusion coefficient ($D$), and of the mean lifetimes of the investigated features are compared with those obtained in this study in Table~\ref{table:compGamma}. Since the effect of atmospheric seeing may be important, it is indicated if the observations are space, ground, or balloon-based. The spatial resolutions of the observations are provided where known. The types of solar regions that were investigated can be deduced from the names of the features. With the exception of the investigations finding super-diffusive motions, discussed earlier, almost all other authors interpreted their results as indicative of normal- or sub-diffusive processes. \citet{Berger1998b} did find a $\gamma=1.34\pm0.06$. However, they interpreted it in terms of normal-diffusion with a slight indication for super-diffusivity, since the area coverage of their network MBPs as a function of time could be well explained by a Gaussian model. \citet{Cadavid1999} found normal-diffusion ($\gamma=1.10\pm0.24$) for timescales longer than $25~\mathrm{min}$ and sub-diffusive MBPs ($\gamma=0.76\pm0.04$) on timescales shorter than $22~\mathrm{min}$, both based on tracking MBPs in a network area.

\subsection{Diffusion coefficient}
\label{sec-diffConclusions-coefficient}

We also determined a mean diffusion coefficient ($D$; the area that a MBP moves across per unit time) of $257\pm32~\mathrm{km}^{2}\mathrm{s}^{-1}$ averaged over all studied MBPs. The diffusion coefficient values reported in the literature and summarised in Table~\ref{table:compGamma} lie between $0.87~\mathrm{km}^{2}\mathrm{s}^{-1}$ and $350~\mathrm{km}^{2}\mathrm{s}^{-1}$. One source of such a large range could be that different values could refer to different features. \citet{Schrijver1996} noted the flux dependence of the diffusivity of flux concentrations, with the smaller concentrations moving faster compared to the larger ones. Tracking small features on relatively short timescales is another source of bias, since the potential effects of larger scales (e.g. supergranular flows) are detected more clearly when measuring for a sufficiently long duration~\citep{Schrijver1996}. By tracking magnetic features in the large FOV of MDI magnetograms, \citet{Hagenaar1999} found that the diffusion coefficient, $D=70-90~\mathrm{km}^{2}\mathrm{s}^{-1}$, determined for intervals of time less than $1.0\times10^{4}~\mathrm{sec}$ is smaller than that measured on timescales longer than $3.0\times10^{4}~\mathrm{sec}$, which gives $220\leq D \leq250~\mathrm{km}^{2}\mathrm{s}^{-1}$. They interpreted this difference (using a model) as the effect of supergranular flow that acts as a negligible drift on short timescales. \citet{Berger1998b} determined $D=60.4\pm10.9~\mathrm{km}^{2}\mathrm{s}^{-1}$ for network MBPs by assuming, for simplicity, Gaussian (normal) diffusion (i.e. $D=sd/2d \tau$), whereas the $\gamma$ they obtained, $\gamma=1.34\pm0.06$, corresponds to the super-diffusive regime. They also found a $D=285~\mathrm{km}^{2}\mathrm{s}^{-1}$ in quiet-Sun internetwork regions by tracking markers (so-called corks) added on frames of image-sequences. \citet{Cadavid1999} discussed the mismatch between different reported diffusion coefficients in the literature due to the assumption of normal-diffusion for all investigated features (i.e. determining $D$ by assuming $\gamma=1$ in Eq.~\ref{equ:diffcoeff2} for simplicity, regardless of actual diffusion indices) in most of the studies to that date. \citet{Roudier2009} estimated $D=430~\mathrm{km}^{2}\mathrm{s}^{-1}$ from Hinode/SOT observations using floating corks on a relatively large scale (larger than $2.5~Mm$; since they used granules to determine the flow). \citet{Chae2008} reported the smallest value of $D=0.87\pm0.08~\mathrm{km}^{2}\mathrm{s}^{-1}$ from Hinode/SOT magnetograms. They estimated $D$ by modelling the change of magnetic field in individual pixels between two frames (taken 10 minutes apart in time), based on solving the equation of magnetic induction. Their unusually small value, however, could likely be caused by the use of individual pixels instead of clearly separated magnetic features in their somewhat different method of diffusivity measurements. Later, \citet{Utz2010} found $D=350\pm20~\mathrm{km}^{2}\mathrm{s}^{-1}$ for internetwork MBPs observed by the same telescope (i.e. Hinode/SOT). They obtained this value in the framework of the normal-diffusion process. \citet{MansoSainz2011} used the same instrument and found $D=195~\mathrm{km}^{2}\mathrm{s}^{-1}$ for footpoints of small-scale internetwork magnetic loops. We speculate that this relatively small value compared to that found in the present work is due to the fact that they considered only freshly emerged small loop footpoints, which may not follow surface flows to the same extent as our magnetic features.

\citet{Cameron2011} described the decay of the magnetic field by turbulent diffusion through 3D radiative MHD simulations. They characterised the decay in terms of diffusion coefficients and found $D$ to lie in the range of $100-340~\mathrm{km}^{2}\mathrm{s}^{-1}$. This range encompasses the mean value of $D=257\pm32~\mathrm{km}^{2}\mathrm{s}^{-1}$ we obtained from the dispersal of {\sc Sunrise} Ca~{\sc ii}~H MBPs.

\subsection{A relationship between diffusion coefficient and timescale}

Furthermore, we found a direct correlation between $D$ and timescale ($\tau$) computed from trajectories of all MBPs (see Fig.~\ref{fig:t_D}). We note that the dependence of $D(\tau)$ on $\gamma$ is not surprising though, since it follows the relationship expressed in Eq.~\ref{equ:diffcoeff2}. The $D(\tau)$ trend we obtained from our super-diffusive MBPs tends to be steeper than those measured by \citet{Lawrence2001} and \citet{Abramenko2011} for internetwork super-diffusive features. \citet{Abramenko2011} compared such a relationship with other types of regions (e.g. network areas, active regions and a coronal hole area) and other diffusion regimes reported in the literature. They showed that such a direct correlation between $D$ and $\tau$ is only observed for the case of super-diffusion ($\gamma>1$). Values of $D$ independent of $\tau$ for $\gamma=1$ \citep{Schrijver1996,Berger1998b,Hagenaar1999} as well as anti-correlations between $D$ and $\tau$ \citep{Berger1998b,Cadavid1999} for $\gamma<1$ are also observed.

\subsection{Migration of MBPs over a supergranule}

As stated earlier, our Monte Carlo simulation describes the motion of our MBPs as the superposition of a directed systematic velocity  $v_{s}$ and a random velocity $v_{r}$. A comparison of the model output with observations of the horizontal motion of the Ca~{\sc ii}~H MBPs allowed us to determine both the random and systematic components as well as the effective coherence time of the random walk flows. We determined the mean values of these velocities to be $v_{r} \approx1.2\pm0.1~\mathrm{km}\, \mathrm{s}^{-1}$ and $v_{s}\approx0.75\pm0.06~\mathrm{km}\, \mathrm{s}^{-1}$. A coherence time of $62\pm3~\mathrm{sec}$ was found for the random velocity.

The random component of the velocity obtained here is consistent with that computed for the rms value of the horizontal transport of the magnetic concentrations in intergranular areas from 3D radiative MHD simulations~\citep{Cameron2011}.

Horizontal flows in supergranules can be measured through the tracking of corks, for example. \citet{Spruit1990} tracked corks over a mesogranule and found a cork velocity of $1.0~\mathrm{km}\, \mathrm{s}^{-1}$ as they move towards the mesogranular boundaries (on timescales of about $10-30~\mathrm{min}$). They showed that the flow speed decreases when approaching the mesogranular boundaries, reaching a minimum of about $0.5~\mathrm{km}\, \mathrm{s}^{-1}$. This range of systematic velocities caused by the combination of granular, meso- and supergranular flows is comparable with that found in the present work. The $v_{s}$ obtained in our investigation is also in good agreement with that of \citet{DelMoro2007}, who found a horizontal flow speed of $0.75\pm0.05~\mathrm{km}\, \mathrm{s}^{-1}$ inside a supergranule via cork tracking. Our determined systematic velocity is, however, slightly larger than the mean horizontal flow velocity of $\approx0.4~\mathrm{km}\, \mathrm{s}^{-1}$ within supergranules, reported by~\citet{Title1989}, \citet{Wang1995a}, and \citet{Hathaway2002} and the $\approx0.3~\mathrm{km}\, \mathrm{s}^{-1}$ reported for mesogranular flows by \citet{Leitzinger2005}.

Recently,~\citet{Orozco2012c} studied the horizontal velocity of both convective flows and internetwork magnetic elements (IMEs) over a SG during a $13$-hour uninterrupted observing campaign with Hinode/NFI. They found that the IMEs are almost at rest at the centre of the SG where they start accelerating radially outward. The IMEs also tend to decelerate while approaching the SG boundaries.

We speculate that a hypothetical long-lived internetwork MBP (that may appear or disappear at any location in the body of a SG) would gently accelerate over the SG's bulk. This is caused by the increasing velocity with radial distance of the supergranular flow profile due to mass conservation assuming a constant upflow over most of the SG's area. The supergranule's flow systematically advects all MBPs within its bulk towards its borders. In addition, granular and mesogranular flows (that act on shorter spatial scales) would impart the MBP with additional velocity components that change as these convection cells evolve. The trajectory of such a MBP is expected to follow a super-diffusive regime, although with a lower $\gamma$ than expected from the SG flow alone, because of the random motions imparted mainly by granular evolution (which for a very long-lived MBP contributes mainly to $v_{r}$) and intergranular turbulence. This MBP would start decelerating when approaching the SG boundary. Once the MBP finds itself in the network region, it is trapped in the sinks (stagnation points) due to inflows from opposite directions, i.e. from neighbouring SGs. The ever-evolving granular flows will keep acting on small spatial scales. The trajectory of the MBP at this stage of its evolution is probably best explained by a normal- and/or sub-diffusion processes.

This scenario may explain why the sub- and normal- diffusive MBPs were almost the only diffusion regimes reported in the literature before $\approx2000$~\citep{Berger1998b,Cadavid1999,Lawrence2001}; older studies most likely concentrated on network areas in which MBPs are easier to observe compared to internetwork regions. For ground-based observations, another effect is also important: the residual aberrations and distortions due to variable seeing introduce an artificial turbulent motion into image time series. The artificial turbulence motion leads to slower growth of $sd$ with time which in turn results in artificially small $\gamma$s (see Eq.~\ref{equ:msd}). It is worth noting that lower spatial resolution observations need not lead to smaller $\gamma$ values and the larger scale flows (e.g. supergranular flows) would be detected more easily compared to smaller scale motions. This effect results in a larger $\gamma$ value.

\subsection{Summary}

To summarise, we characterised the motions of {\sc Sunrise} Ca~{\sc ii}~H MBPs by turbulent diffusion theory. The MBPs (mean lifetime $\approx8~\mathrm{min}$) were observed in seeing-free high-resolution image sequences in an internetwork area of the quiet Sun. A mean diffusion index of $\gamma=1.69\pm0.08$ and a mean diffusion coefficient of $D=257\pm32~\mathrm{km}^{2}\mathrm{s}^{-1}$ were obtained. The $\gamma$ corresponds to super-diffusion which describes the MBPs as features whose squared displacement ($sd$) from the first observed location grows faster than linearly with time. It is, to our knowledge, the largest $\gamma$ value for MBPs reported in the literature so far. The parameter $D$ lies within the range of decay rate of the magnetic field from MHD simulations and is among the largest $D$ values obtained for small-scale magnetic features found in the literature. We found that $D$ increases as the timescale increases, but generally lies in the range of those obtained by other investigations for larger spatial extent and longer durations. The migration of relatively short-lived features such as MBPs is composed of a superposition of random motions due to granular evolution and intergranular turbulence and systematic motions due to more steady granular evolution, mesogranular, and supergranular flows.

\begin{acknowledgements}

The German contribution to {\sc Sunrise} is funded by the Bundesministerium f\"{u}r Wirtschaft und Technologie through the Deutsches Zentrum f\"{u}r Luft- und Raumfahrt e.V. (DLR), Grant No. 50 OU 0401, and by the Innovationsfond of the President of the Max Planck Society (MPG). The Spanish contribution has been funded by the Spanish MICINN under projects ESP2006-13030-C06 and AYA2009-14105-C06 (including European FEDER funds). The HAO contribution was partly funded through NASA grant NNX08AH38G.

\end{acknowledgements}

\bibliographystyle{aa} 
\bibliography{ref_diffusivity} 

\end{document}